\documentclass{sig-alternate}
\pdfoutput=1
\usepackage{enumerate}
\usepackage{makeidx}  % allows for indexgeneration
\usepackage{graphicx}
\usepackage{subfigure}
\usepackage[english]{babel}
\begin{document}
%
% --- Author Metadata here ---
\conferenceinfo{Multimedia}{}
\CopyrightYear{2013}
\crdata{}
\clubpenalty=10000
\widowpenalty = 10000
% --- End of Author Metadata ---

\title{Learning Subclass Representations for Visually-varied Image Classification}

\author{
% 1st. author
\alignauthor
Xinchao Li, Peng Xu, Yue Shi, Martha Larson, Alan Hanjalic\\
       \affaddr{Multimedia Information Retrieval Lab, Delft University of Technology}\\
       \affaddr{Delft, The Netherlands}\\
       \email{\{x.li-3, p.xu, y.shi, m.a.larson,a.hanjalic\}@tudelft.nl}
}
% Just remember to make sure that the TOTAL number of authors
% is the number that will appear on the first page PLUS the
% number that will appear in the \additionalauthors section.

\maketitle

\begin{abstract}
%Image classification that uses exclusively the visual content of images holds the
%promise of effectively facilitating users' access to media collections.
In this paper, we present a subclass-representation approach that predicts the probability of a
social image belonging to one particular class. We explore the co-occurrence of user-contributed tags to find subclasses with a strong
connection to the top level class. We then project each image on to the resulting subclass space to
generate a subclass representation for the image. The novelty of the
approach is that subclass representations make use of not only the content of the photos themselves, but also information on the co-occurrence of their tags, which determines membership in both subclasses and top-level classes.
The novelty is also that the images are classified into smaller classes, which
have a chance of being more visually stable and easier to model. These
subclasses are used as a latent space and images are represented in this space
by their probability of relatedness to all of the subclasses. In contrast to approaches
directly modeling each top-level class based on the image content, the proposed method
can exploit more information for visually diverse classes. The approach is
evaluated on a set of $2$ million photos with 10 classes, released by the
Multimedia 2013 Yahoo! Large-scale Flickr-tag Image Classification Grand
Challenge. Experiments show that the proposed system delivers sound
performance for visually diverse classes compared with methods that directly
model top classes.
\end{abstract}

% A category with the (minimum) three required fields


\section{Introduction}
\label{sec:intro}

%Content-based image classification and retrieval have been studied extensively in the past two decades~\cite{Smeulders2000,Zhang2007a}, and are used to train predictive models for inferring image classes or ranking images for given queries, based on the content features (visual features) of images. On another side, the explosive growth of social media, together with the ubiquitousness of social tagging function, have enriched the quantity of labeled (tagged) images on the Web to an extremely large extent, e.g., Flickr\footnote{http://www.flickr.com/}. With such enormous amount of social data being annotated by users, it becomes interesting to develop techniques that can learn classifiers when the number of images per class is very large.
%Although social tagging has been recognized as a great contribution to image classification and retrieval, the ambiguity and the visually irrelevance of user-generated tags have become a new bottleneck for image retrieval systems.
%For example, a lot of images in Flickr are tagged with ``nature'', while these images have a huge variation in their visual features, e.g., some images are about mountains or lakes, and some about DNA sequences.
%As a result, classifying or retrieving images regarding such tags (classes) solely based on visual features becomes an extremely challenging task.
This paper describes the approach that we developed to address the Yahoo! Large-scale Flickr-tag Image Classification Grand Challenge. This challenge is formulated:

\begin{figure}[t]
  \centering
  \scalebox{0.45}{\includegraphics[width=\textwidth]{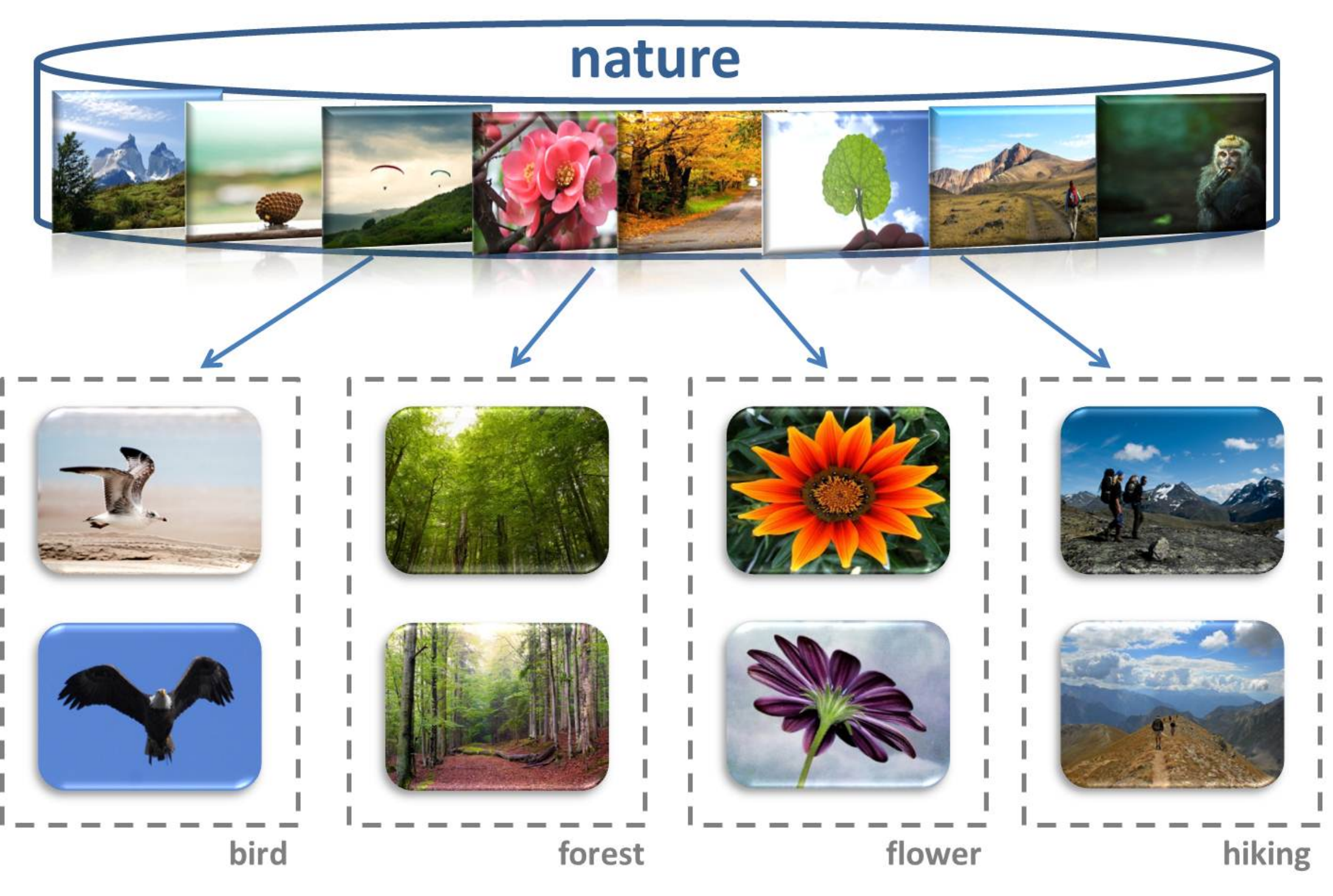}}
  \caption{Example illustration of the subclass representation}\label{fig:examp}
  \vspace{-0.75cm}
\end{figure}

\emph{Given a training set of images together with their metadata, and a class
label corresponding to each training image, to build a ranking model that for
each class label ranks images in a test set (without metadata) as accurate as
possible.}

We emphasize three particular characteristics of this challenge. First, the ground truth for both the training and testing set is user generate tags. As with any other tagging problem, these tags are expected to be noisy, incomplete and subjective. The ranking models should have the ability to learn useful information with difficult labels. Second, for this particular task, the 10 target tags are selected from tags most frequently used by Flickr users. The images from these classes (which we refer to as `top-level classes') show a broad variation in terms of visual representations. Thus, the ranking model should be based on a representation that can model classes with high intra-class variations. Third, the level of visual diversity varies between the given 10 target tags. For example, images related to tag ``sky'' or ``beach'' are expected to be more visually consistent than images related to ``2012'' or ``nature''. In view of this, the ranking model should have the ability to deal with different abstraction levels adaptively.

%in the test set, images are provided without metadata. In other words, in the test set, only the visual information can be used for ranking the images for each query. Thus, the ranking model that is expected for this challenge should be a function of variables that can be derived from visual information of images. Second, the goal is to rank images for each class label rather than to only predict the class label for each image. In other words, the class labels are used as queries to retrieve images, which are expected to be ranked in a descending order of their relevance to the corresponding query. More specifically, the main evaluation metric for this challenge is set to be average precession (AP), which is a top-biased ranking metric. Thus, the ranking model is expected to place correctly classified images for each class label as early as possible. Third, the class labels in the challenge are chosen from social tags. In other words, the class labels can be regarded as a special set of tags, which may be hardly reflected by the visual information or have a broad variation in terms of visual representations. Thus, the ranking model is expected to be particularly effective for retrieving/classifying visually-varied images for ambiguous class labels.

\begin{figure}[t]
  \centering
  \scalebox{0.45}{\includegraphics[width=\textwidth]{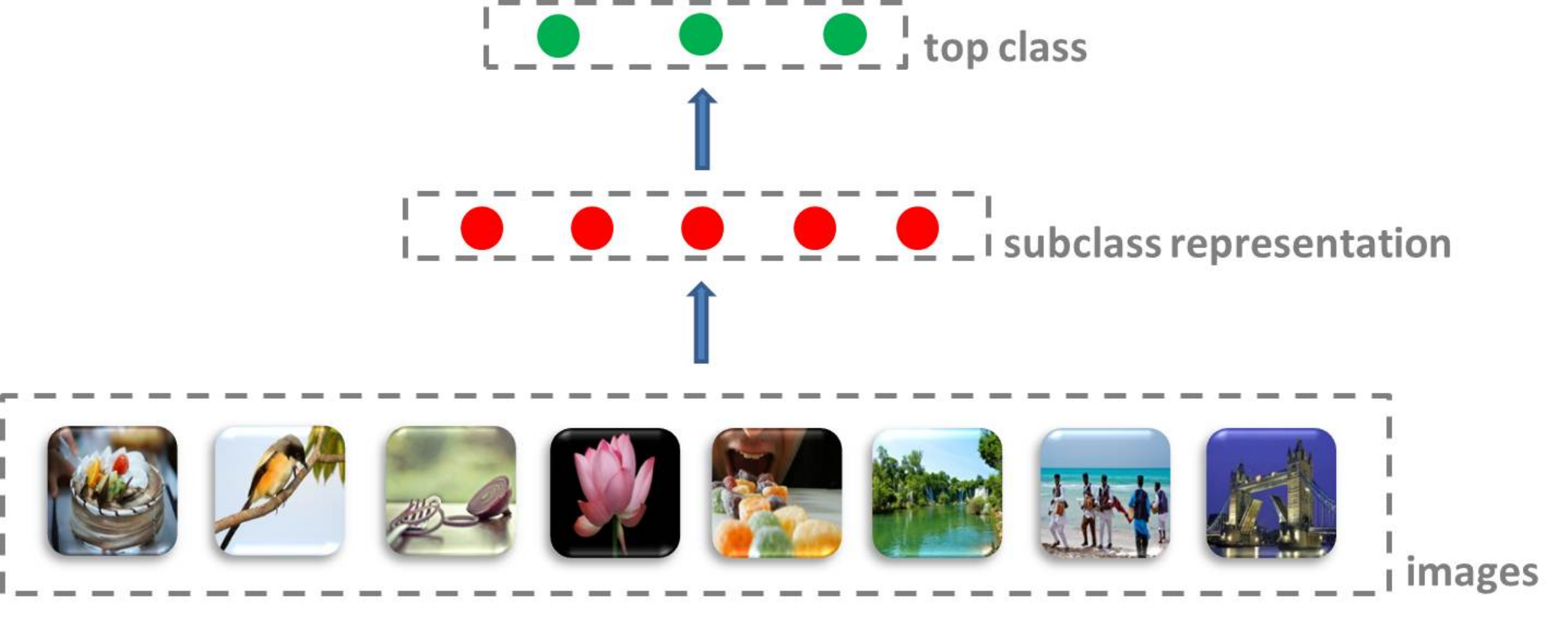}}
  \caption{Overview of the propose approach. First subclasses are discovered, then top-level classes are predicted using subclass representations.}\label{fig:scheme}
\end{figure}

In view of the above, we propose in this paper a new method for content-based image classification/tagging based on subclass
representation. As mentioned above, the image classes in this challenge are user-generated tags. Since social images are usually annotated by more than one tag, the co-occurrence of different tags serves as a valuable information source that can be exploited for elaborating the tags (classes) which are by themselves highly visually-varied. As shown in the example in Fig.~\ref{fig:examp}, the image class ``nature'' can cover an extremely large range of visual representations. However, the subclasses of it, such as ``flower'', ``bird'', ``forest'', are much more homogeneous in terms of the visual information. For this reason, we first discover the subclasses, which are tags frequently and exclusively co-occurring with the top-level class labels, for representing each image. Then, we train a binary classifier for each selected subclass. For a given image, the confidence scores of all subclass classifiers are concatenated to produce a high level representation.
Thus, the representation is developed to learn models of the top-level classes. This is illustrated by Fig.~\ref{fig:scheme}.
The final results are predicted by the ranking model based on the the learned subclass representation.

The contribution of this paper lies in the following aspects:
\begin{itemize}
\item This method uses a co-occurrence based method to discover subclasses. Compared with semantic ontology based subclass generation methods~\cite{DengJ2009} or using predefined concepts as subclasses~\cite{Torresani2010}, the proposed subclass representation is expected to be more discriminative in terms of predicting the target classes.
\item The proposed method uses confidence score instead of binary decision of the subclass classifiers as the high level features. This strategy can develop useful representations even if the performance of subclass classifier is not reliable.
\end{itemize}

The remainder of the paper is organized as follows: in Section~\ref{sec:system}, we present the details of the proposed method. The experimental framework and results are presented in Section~\ref{sec:expFrame} and Section~\ref{sec:expResult}. Then, in Section~\ref{sec:rw} we discuss previous research contributions that are related to our approach proposed in this paper. Finally, Section~\ref{sec:conclude} summarizes our contributions and discusses future work.

\section{Learning Subclass Representation}  \label{sec:system}
\subsection{Mining Subclasses}
As discussed above, subclasses of one class (the target class) are expected to
be strongly connected with the target class and, moreover, relatively more
stably reflected in visual features than the target class.
To define such subclasses, we exploited the tags annotating the images. We first generate a co-occurrence matrix between photos'
tags and their top-level classes, and measure each tag's connection to one class by its
distinctive score, defined as:

\begin{equation}
S_{ij}  = \frac{{C_{ij} }}{{\sum\limits_j {C_{ij} } }}
\end{equation}
where $C_{ij}$ is the number of co-occurrence of the $i$-th tag and the $j$-th
top level class. Note that in the setting of Yahoo! Challenge, the predefined to-level classes are
also chosen from the user-contributed tags.

The selected subclasses for class-$j$ can be then defined as the tags that have
their distinctive scores above a pre-defined threshold, as shown below:
\begin{equation}
Tsel_j  = \{ t_{i} |S_{ij}  > Thr\_distin,t_{i}  \in T\}
\label{thrhod}
\end{equation}
where $t_i$ denotes the $i$-th tag, and $T$ denotes the set of all tags.
Note that some tags may be only assigned to a very small number of images
in one class. For those tags, the limited number of training images prevents
them from being effective subclasses. Taking this into account, we
further rank all selected tags, $t_i$, in $Tsel_j $ by the number of photos in
class-$j$ that are tagged with $t_i$.

\subsection{Subclass Representation}
To generate a subclass-based representation for an image, we first use the
images tagged with the subclasses to train models, i.e., Support Vector Machines
(SVMs), for classifying subclasses, and then, use the confidence scores for
predicting each subclass as the new representation for the image. In this sense, as illustrated in Fig.~\ref{fig:scheme},
the image features can be treated as the fist level representation for an image,
while the confidence scores of subclasses are the high level representation. Based on the subclass representations, we further learn the model that characterize the connection between these representations and the top-level class.

\section{Experimental Framework} \label{sec:expFrame}
\subsection{Dataset}
To verify the performance of the proposed approach, we carry out our experiments
on a dataset of photos released by the Multimedia 2013 Yahoo! Large-scale
Flickr-tag Image Classification Grand Challenge. The dataset contains 2 million
Flickr photos with 10 classes, i.e., 150K training and 50K test images per class. The class
labels are amongst the top tags annotated by the Flickr users.
%As discussed earlier, most of the 10 classes, as indicated in Fig.~\ref{fig:apSubClass}, are
%highly abstract and visually-varied.
Since the release does not includes the tags associated with the photos, we re-crawled the photos' tags using the photo ID provided in the metadata.

To develop our system, the training dataset is randomly divided into three parts
with the ratio 4:3:1. The first is for training models for
predicting subclasses based on image features, the second is for training models
for target classes based on confidence scores from the learned subclass models,
and the last is for validation and parameter selection. The test data from the grand challenge is used to evaluate the proposed system.

\subsection{Multi-class Classification}
To model the connection between image feature and subclass, and the connection between subclass representation and top class, we choose an SVM-based approach.
For the purpose of classification for multiple classes, we apply one-against-one
training strategy, which is reported to have better training time efficiency and
prediction accuracy compared with other multi-class support vector machines,
e.g., one-against-all~\cite{SVM_1vs1}. To generate probability estimation from
the SVM model, we apply the algorithm proposed in \cite{SVM_prob} and modified
its original implementation in LibSVM~\cite{LibSVM} to make it suitable
for distributed computing on a Hadoop-based distributed server.

\subsection{Baselines}
We compare our subclass-representation-based approach, denoted in the following
as $SVM\_SubClassProb$, to two other approaches that closely relate to our
approach, as listed below.
\begin{itemize}
  \item $SVM\_VisFeat$: Directly model target the 10 top-level classes based on
  image features.
  \item $SVM\_ClassProb$: Project image features on to top-level 10-classes space.
\end{itemize}

\section{Results} \label{sec:expResult}

\subsection{Learning Subclass Models}
\begin{figure}[htb!]
  \centering
  \scalebox{0.43}{\includegraphics[width=\textwidth]{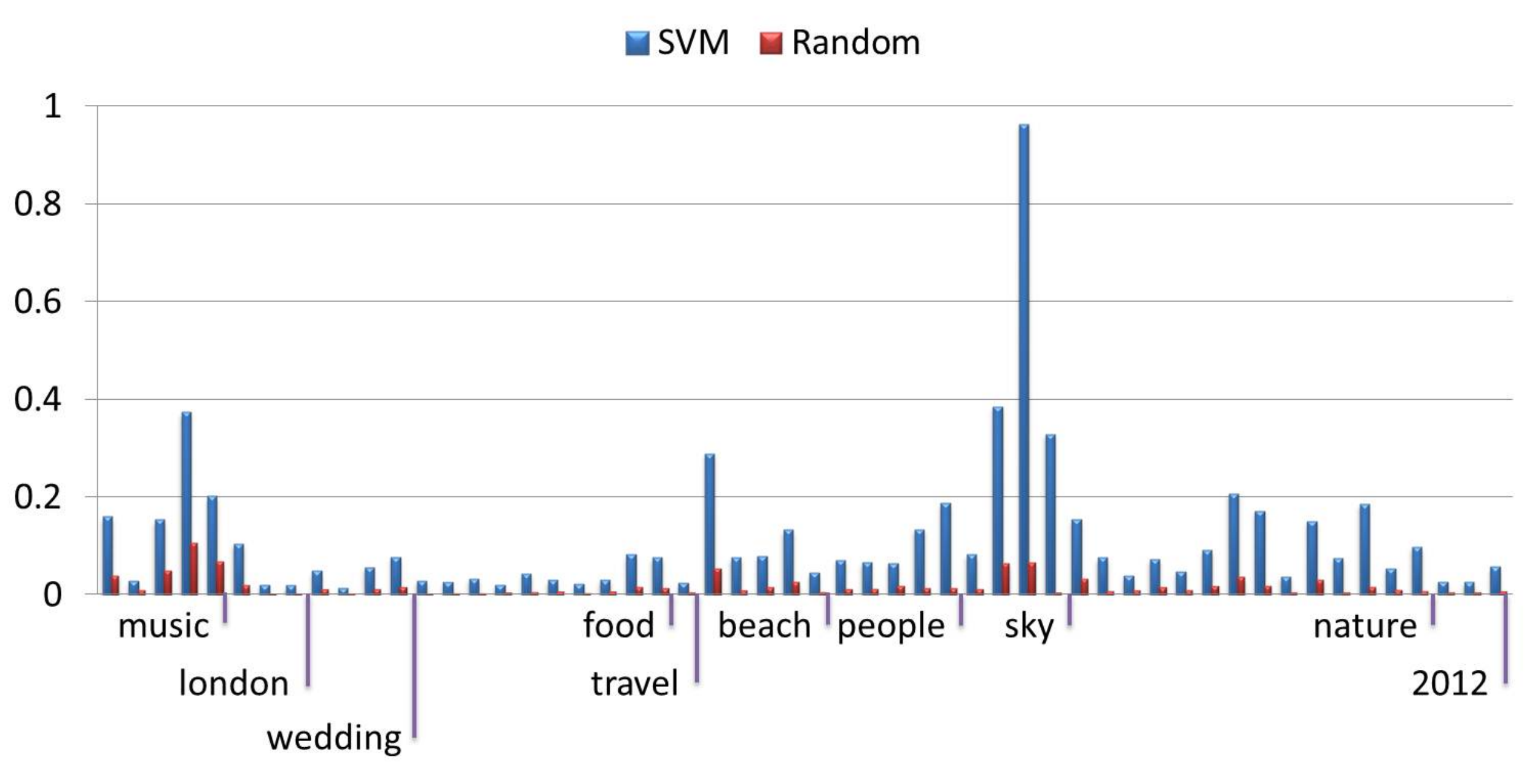}}
  \caption{AP for 54 subclasses, labels on the horizontal axis indicate subclass association with the top-level classes.}\label{fig:apSubClass}
\end{figure}

To learn the subclasses using the co-occurrence matrix between photos' tags and
their classes, the $Thr\_distin$ in Eq.~\eqref{thrhod} is set to $0.6$, and then
we manually select the subclasses in the top $10$ tags of each class, which
results in a total of 54 subclasses. As some classes may contains few distinctive tags
compared with other classes, they may have few subclasses, i.e., class
``travel'' only contain 1 subclass. In contrast, some classes may contain more distinctive tags, i.e., 14 subclasses for class
``nature''. To train these subclass models for projecting images on to the subclass space, we use a maximum 10k images per subclass as training data. Note that some subclasses may contain less than 10k images. The
performance on the validation set, Average Precision (AP) for these subclasses models
are illustrated in Fig.~\ref{fig:apSubClass}.

\subsection{Classification Results}
\begin{figure}[htb!]
  \centering
  \scalebox{0.45}{\includegraphics[width=\textwidth]{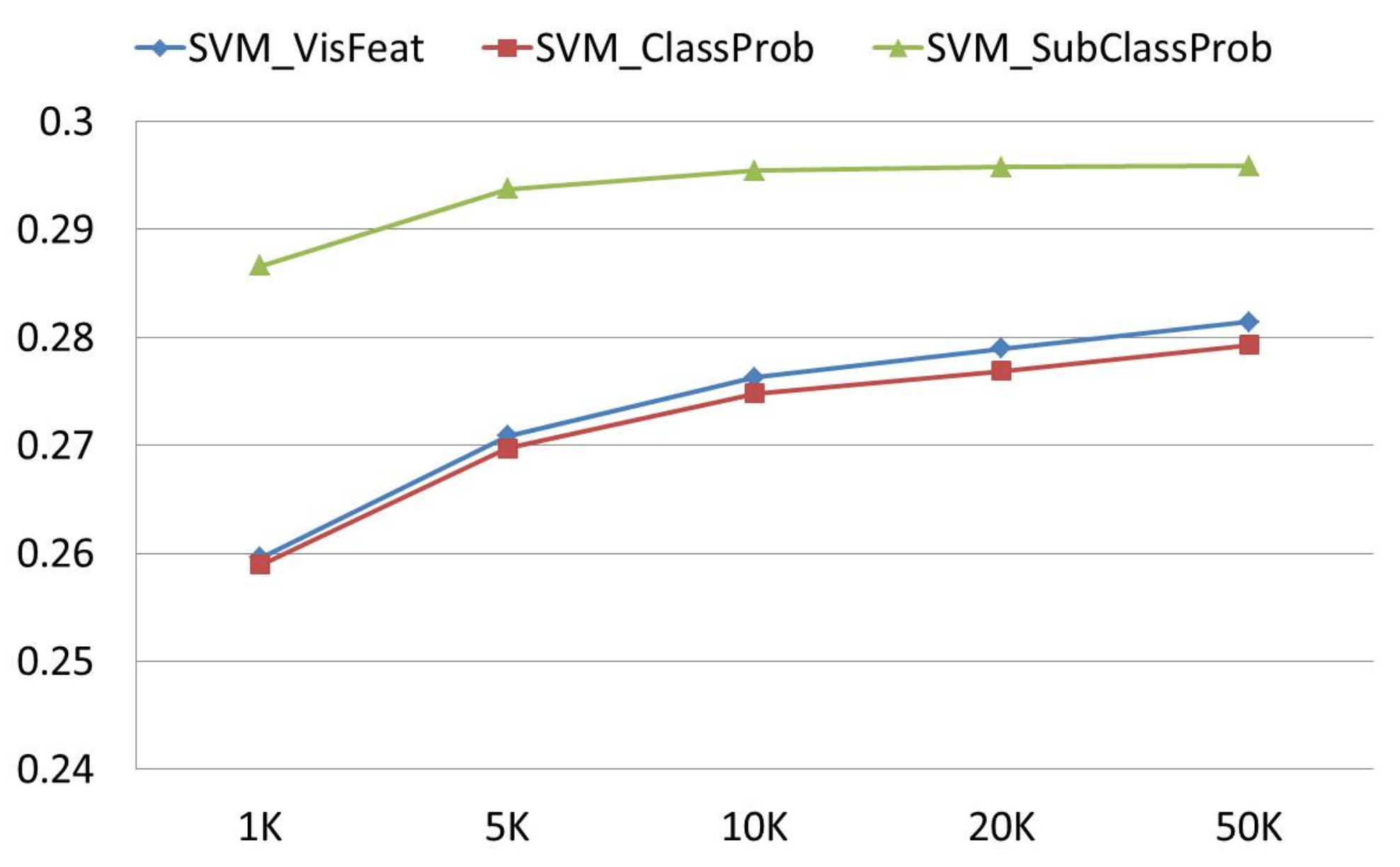}}
  \caption{MAP with different amounts of training data.}\label{fig:MAP}
\end{figure}

Fig.\ref{fig:MAP} illustrates the performance in terms of mean average precision
(MAP), across all 10 top-level classes, for the proposed approach and the
baselines with respect to different training data scales.
Overall, $SVM\_SubClassProb$ performs better than $SVM\_VisFeat$ and
$SVM\_ClassProb$ in different training data scale, with an averaged gain $8\%$
for $SVM\_VisFeat$ and $7\%$ for $SVM\_ClassProb$.
In addition, the $SVM\_SubClassProb$ already achieves good performance compared to others in case of a small training data scale, e.g., 1k per class. Along with the increasing amounts of training data, the performance for $SVM\_SubClassProb$ levels out. This is due to the fact that many subclasses do not contain enough training photos, e.g., $83\%$ of them contains photos which are less than 10k.

\begin{figure}[htb!]
  \centering
  \scalebox{0.4}{\includegraphics[width=\textwidth]{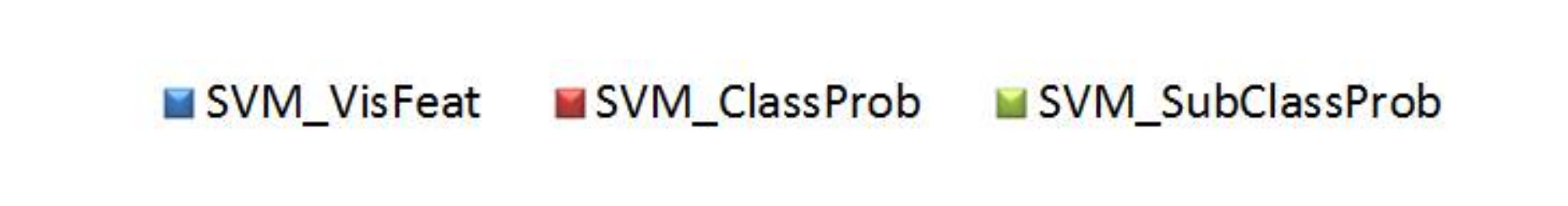}}
  \subfigure[music] {\scalebox{0.23}{\includegraphics[width=\textwidth]{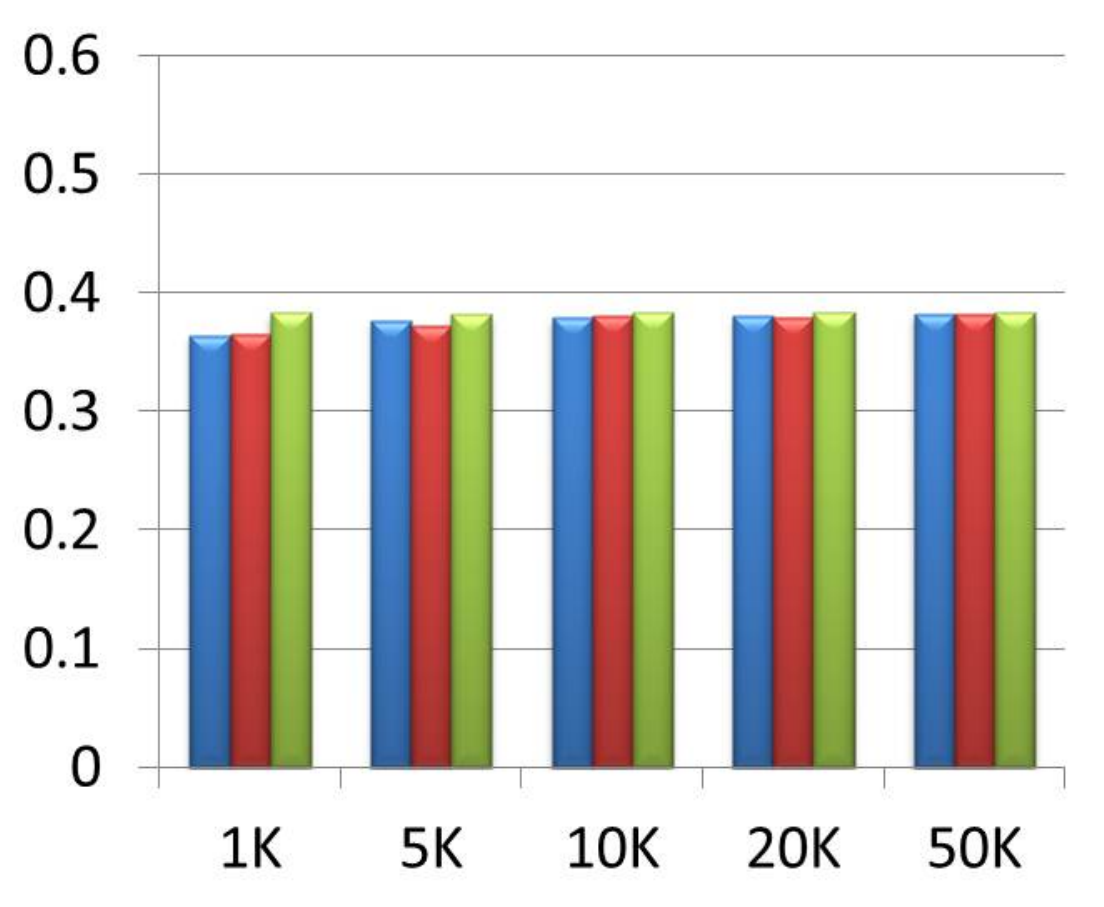}}}
  \subfigure[london] {\scalebox{0.23}{\includegraphics[width=\textwidth]{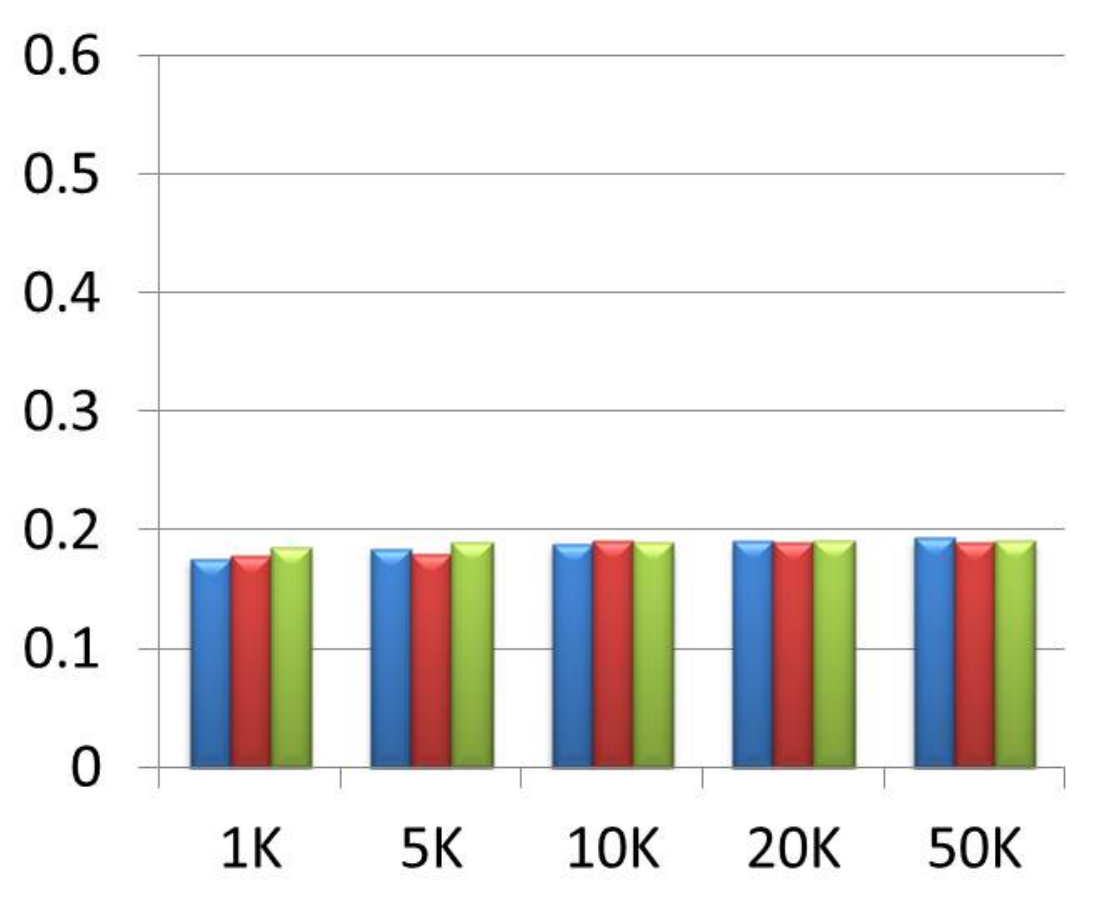}}}
  \subfigure[wedding] {\scalebox{0.23}{\includegraphics[width=\textwidth]{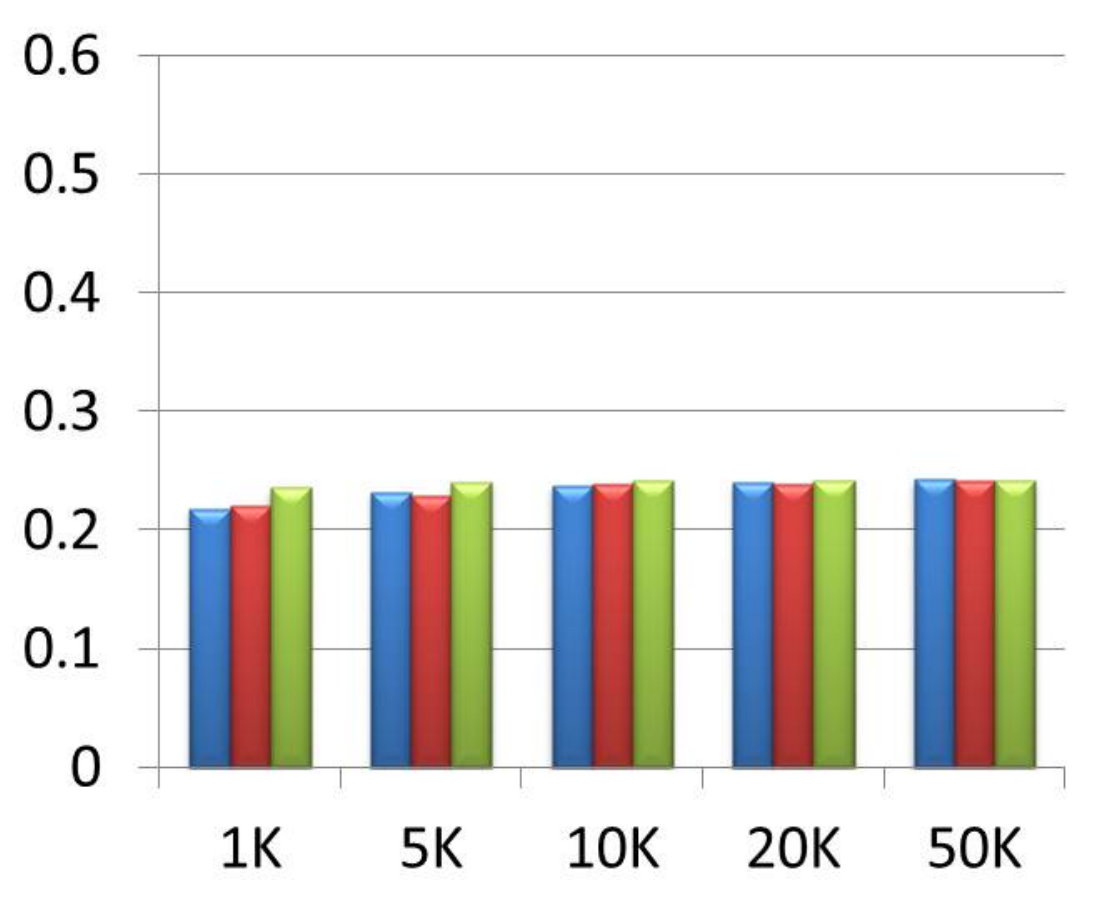}}}
  \subfigure[food] {\scalebox{0.23}{\includegraphics[width=\textwidth]{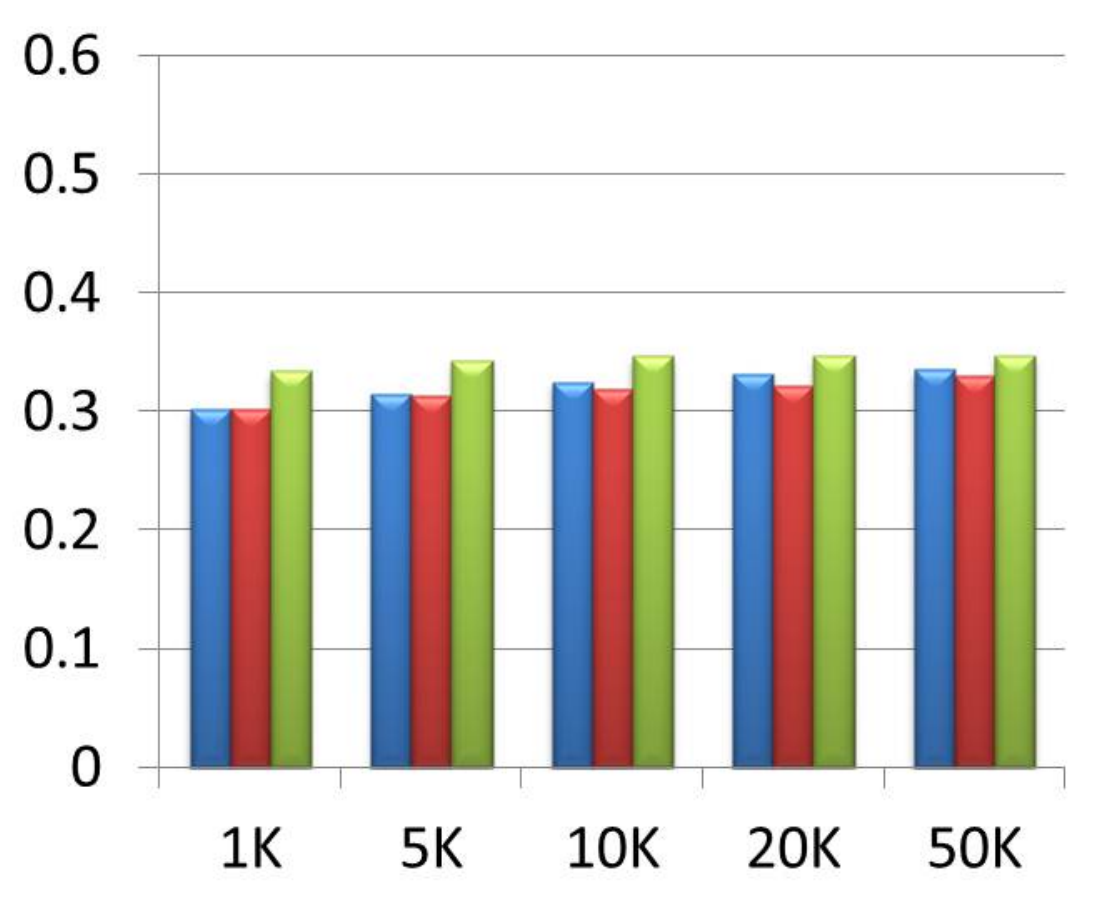}}}
  \subfigure[travel] {\scalebox{0.23}{\includegraphics[width=\textwidth]{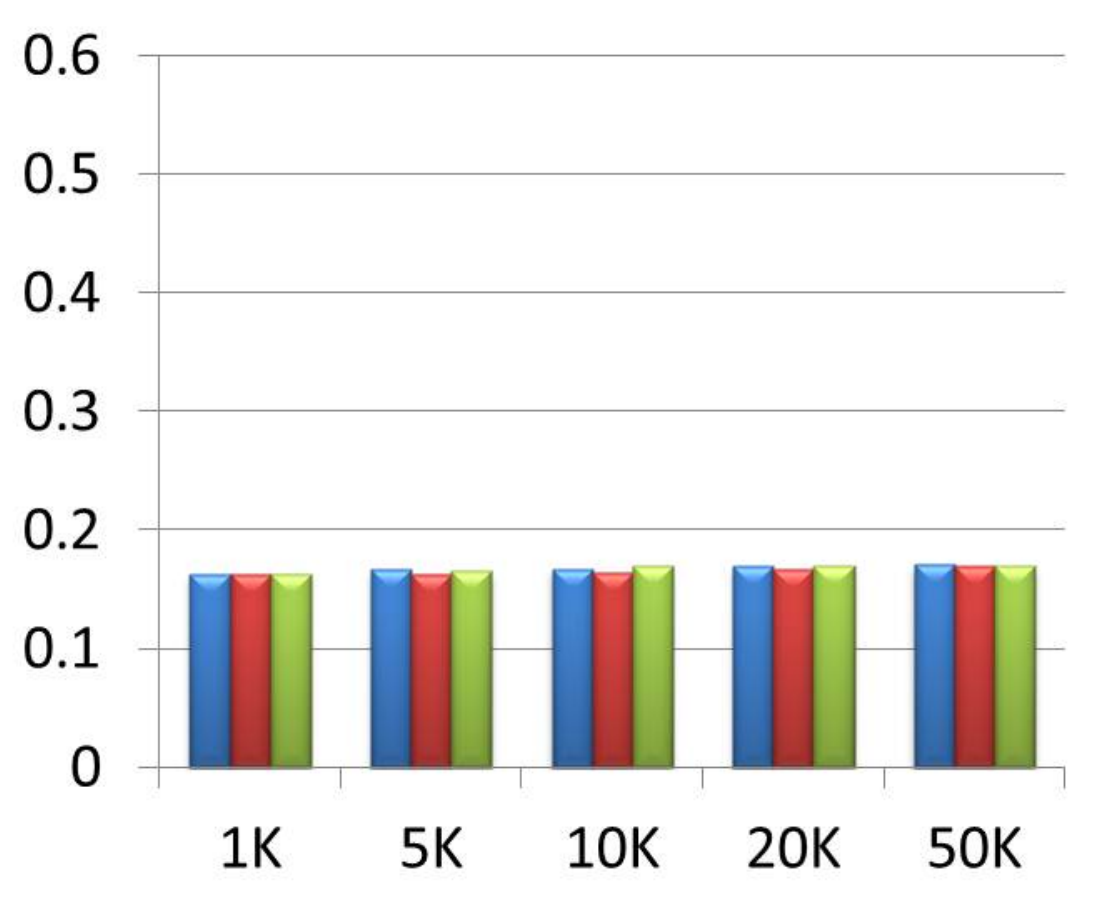}}}
  \subfigure[beach] {\scalebox{0.23}{\includegraphics[width=\textwidth]{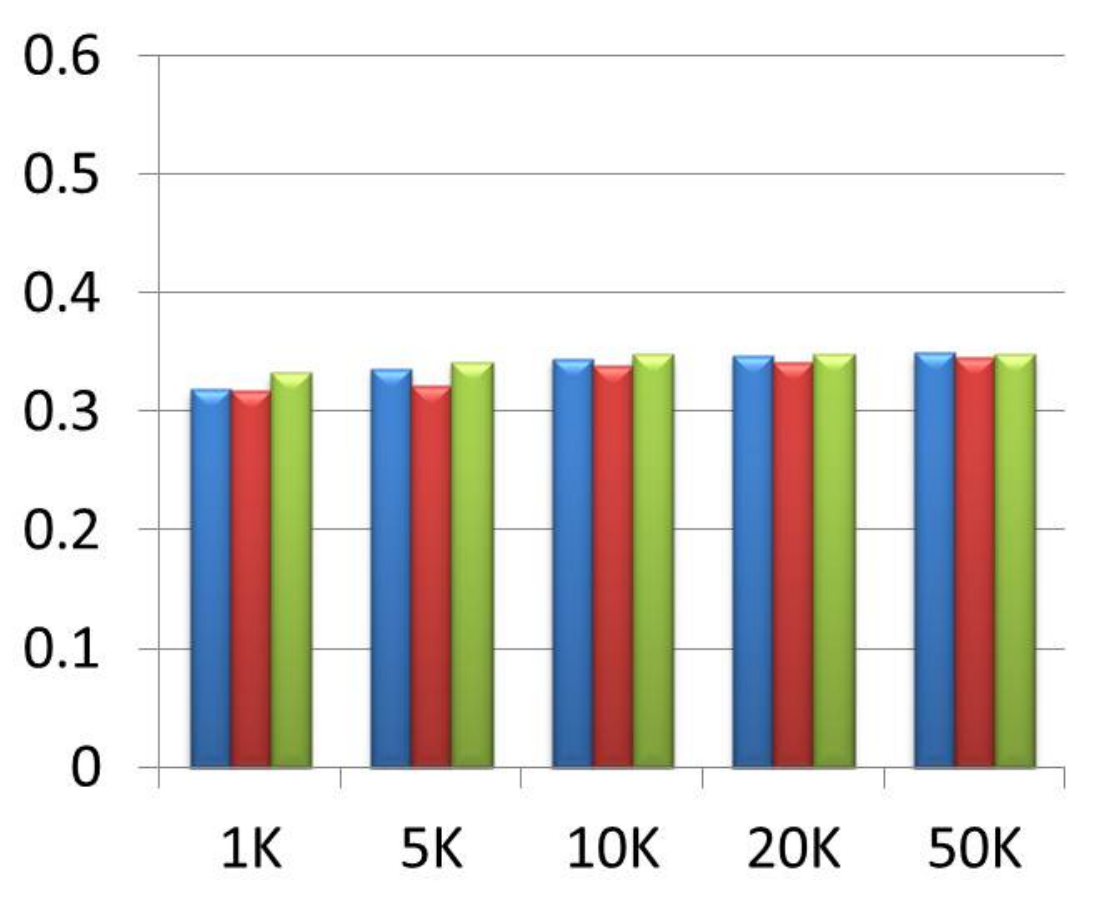}}}
  \subfigure[people] {\scalebox{0.23}{\includegraphics[width=\textwidth]{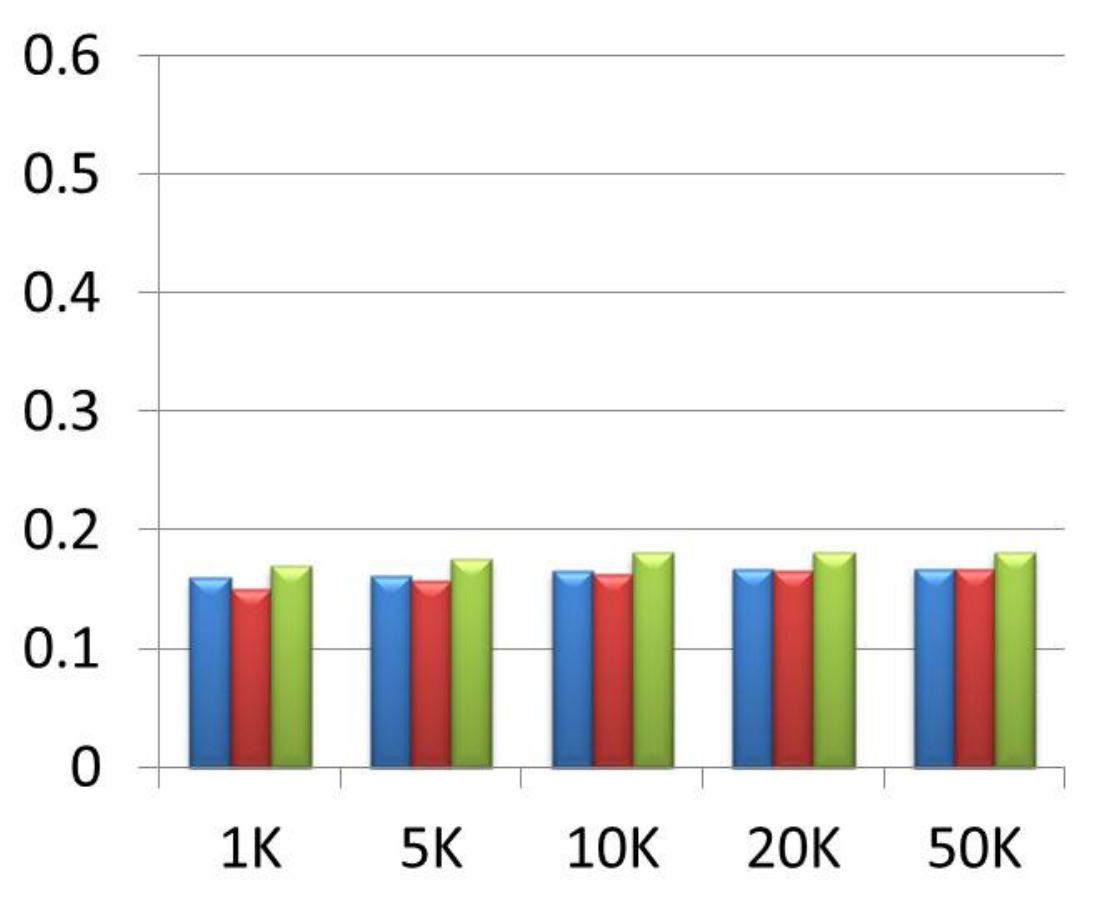}}}
  \subfigure[sky] {\scalebox{0.23}{\includegraphics[width=\textwidth]{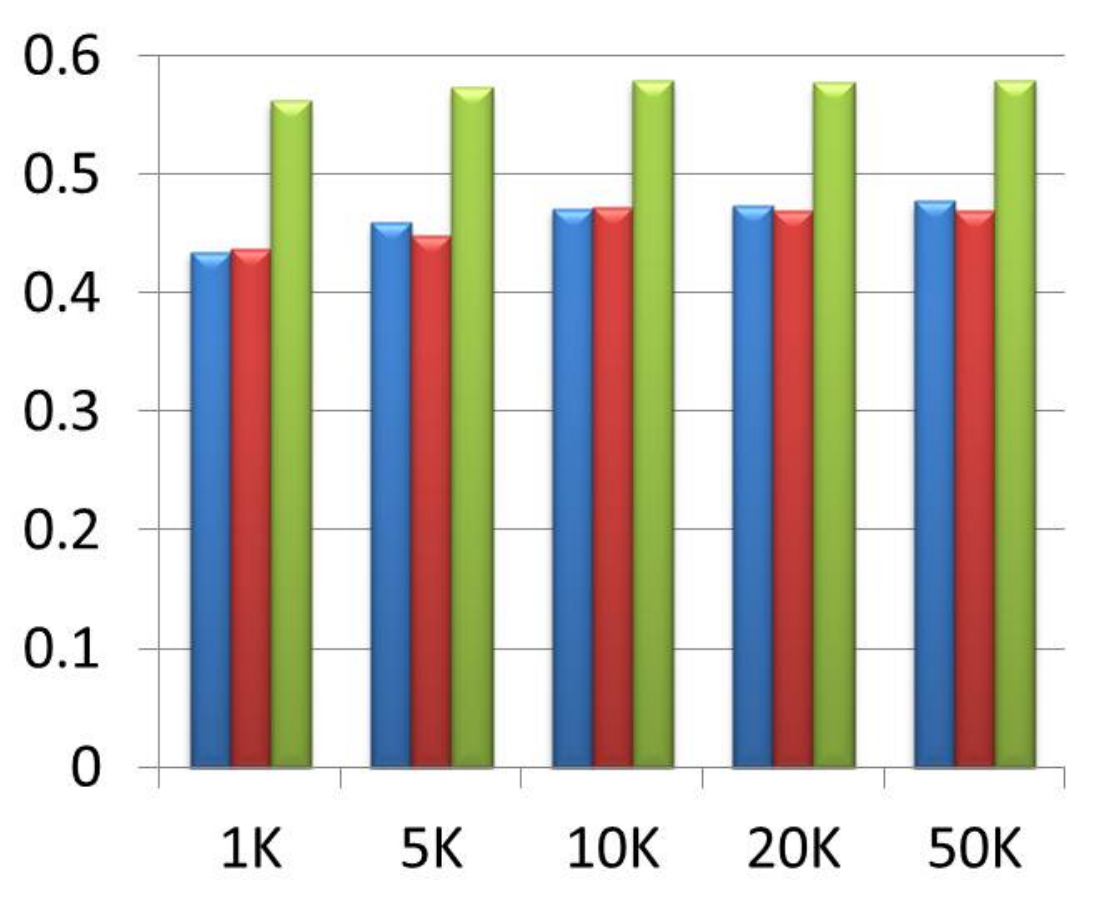}}}
  \subfigure[nature] {\scalebox{0.23}{\includegraphics[width=\textwidth]{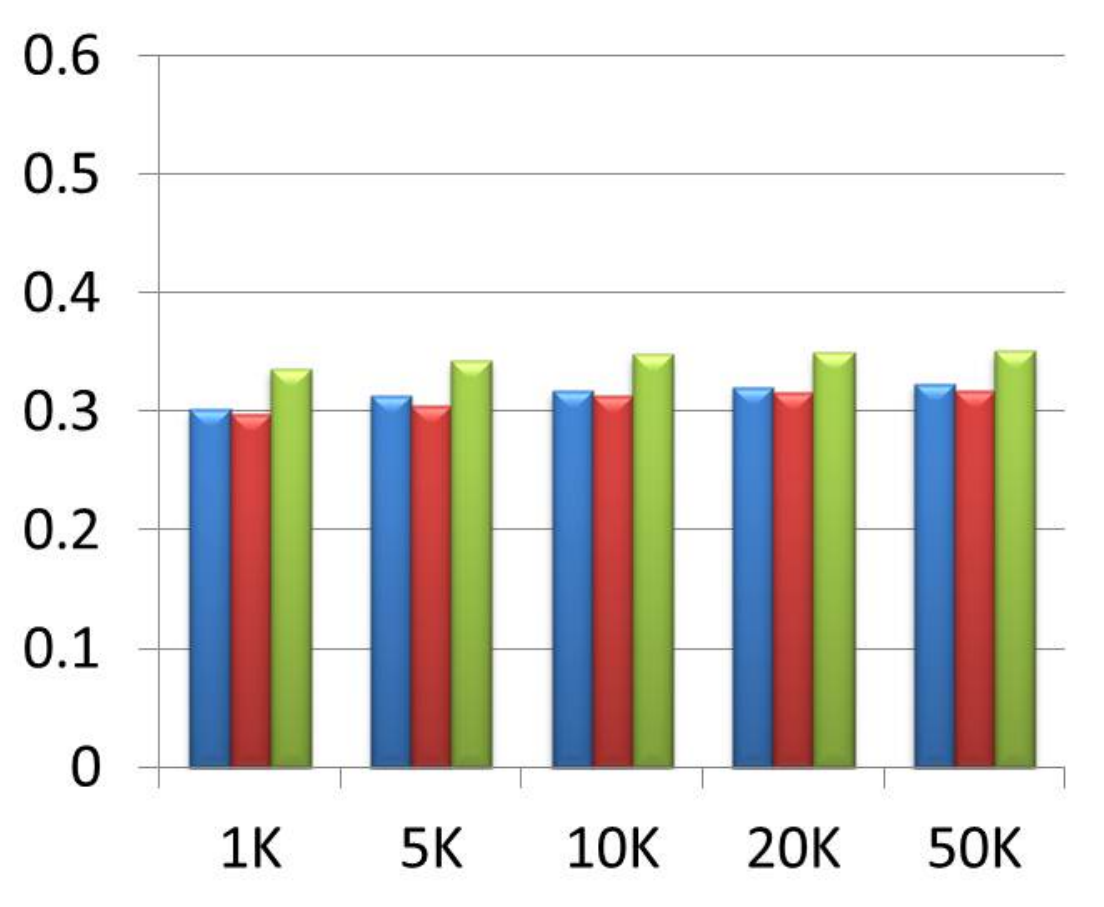}}}
  \subfigure[2012] {\scalebox{0.23}{\includegraphics[width=\textwidth]{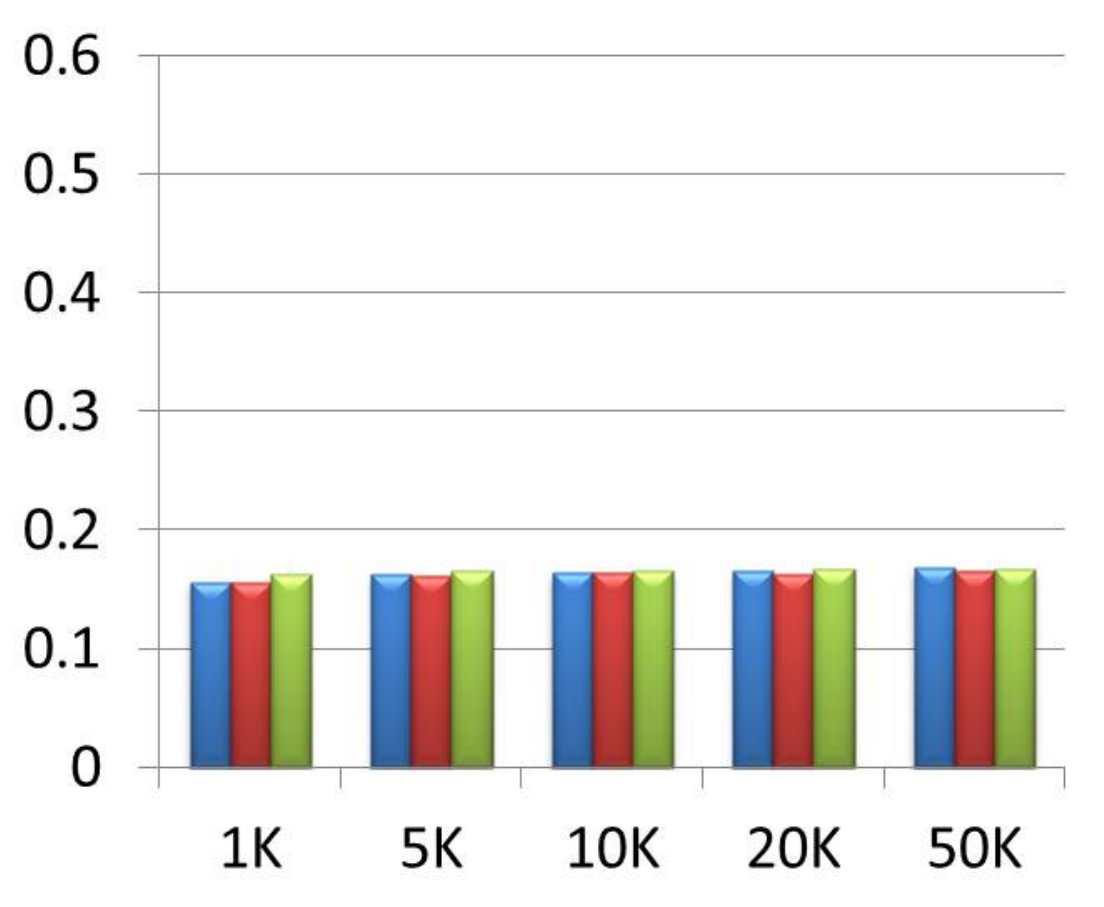}}}
  \vspace{-0.3cm}
  \caption{AP for 10 classes on different training data scale.}\label{fig:ap10Class}
  \vspace{-0.4cm}
\end{figure}

Fig.~\ref{fig:ap10Class} further breaks down the performance in terms of AP over
different classes with respect to different training data scales. As can be
seen, for classes ``food'', ``people'',``sky'',``nature'', $SVM\_SubClassProb$
gains more improvements compared to other classes. This is due to the
fact that, for these classes, they own more subclasses compare to other classes, i.e., there are more
distinctive tags in these classes. Also, some classes, ``sky'',``people'',
contains visually highly consistent subclasses, as illustrated in
Fig.~\ref{fig:apSubClass}, which give a strong support for top classes. This is especially obvious for class ``sky'', in which there is a subclass that has a very strong visual consistency, providing a good support for the top class. Interestingly, for the class ``food'', which owns many subclasses, and each of them has a relative low AP, however, this class still has reliable performance. We conjecture that these subclass classifiers are providing useful discriminative information in form of probabilities that they yield with respect to non-relevant subclasses.

\vspace{-0.2cm}
\section{Related Work} \label{sec:rw}

%The basic image tagging approaches can be categorized into two paradigms, i.e., model-based methods and search based methods.
Our work is closely related to method used for predicting tags with high intra-class variation, in particular, sub-category based methods or methods based on learning high level representations are often explored. We discuss these methods here in turn.
%The basic assumption of the search based image tagging methods is that visually similar images tend to share the same tags. It has been shown that reasonable annotation results can be achieved by accumulating tag information from the visually neighbors of a given image~\cite{Torralba2008, LiX2009}. The main disadvantage of search based methods is that they require more memory to store all the samples, and are expected to be less efficient in terms of prediction time. For large-scale image tagging tasks, the most popular model based image tagging method is to train a binary classifier for each tag~\cite{Sanchez2011, LinY2011, Perronnin2012}. The problem of these methods is that the images of some tags may be too divers to train a reliable model.

Generating sub-categories has been considered as an effective method to deal with classification problems where intra-class variation is high. The ImageNet~\cite{DengJ2009} organizes image dataset with labels corresponding to a semantic hierarchy. This method is able to build comprehensive ontology for large scale dataset. However, for a particular dataset, the sub-categories generated by data driven strategies are expected to be more discriminative. \cite{YangWeilong2011} exploits co-watch information to learn latent sub-tags for video tag prediction. \cite{LiL2010} proposes to discover the image hierarchy by using both visual and tag information.
Our method generates category-specific subclasses by exploring image/tag co-occurrence, and trains classifiers for each subclass-tag. These subclasses based models are expected to be discriminative in terms of estimating the target tags, corresponding to top-level classes.

Learning higher level representation is adopted when the low-level image features are not discriminative enough for the purpose of classification ~\cite{LiL2010ObjectBank, Torresani2010}. For the supervised representation learning methods, a predefined set of models are trained based on image features. The output of these models is considered as the high level representations for predicting image categories. Recently, deep neural networks have been used on unsupervised learning image representations with large scale image dataset~\cite{Lee2009, LeQ2012}. This representation has achieved promising results on different classification and tagging tasks~\cite{Krizhevsky2012}. Our methods used the output of trained subclass classifiers as the high level features. This structure can be easily extend to deeper levels by finding discriminative tags for the subclasses.
\vspace{-0.2cm}

\section{Conclusion} \label{sec:conclude}
We have presented a subclass-representation approach to the task of
retrieving/ranking large scale social images to one particular class solely
based on visual content.
The main contribution of the approach is that by projecting the image feature
representation on to a subclass space generated by exploiting the co-occurrence
information of user-contributed tags, it makes use not only of the content of the photos themselves, but also of information concerning the co-occurrence of the photo's tags with tags corresponding to top-level classes.
%The proposed approach gives the potential of helping classes with high visual
%diversity when the it includes subclasses that are visually homogenous enough to
%be well modeled by content features.
%Future work will include systematical investigation of the subclass mining, not
%only explore the co-occurrence information of user-contributed tags to find
%subclass, but also considering the visual stability within each subclass.
%
% ---- Bibliography ----
%
\vspace{-0.2cm}
\bibliographystyle{abbrv}

\bibliography{xinchaoBib}

%\vspace{2mm}
%\scriptsize{ \bibliography{xinchaoBib} }

%\vspace{2mm}
%\small{ \bibliography{xinchaoBib} }

 % sigproc.bib is the name of the Bibliography in this case

\end{document}